\documentclass[10pt, conference, compsocconf]{IEEEtran}

\usepackage[pdftex]{graphicx}
\DeclareGraphicsExtensions{.pdf,.jpeg,.png}

\usepackage[cmex10]{amsmath}

\usepackage[english]{babel}
\usepackage[T1]{fontenc}
\usepackage{inputenc}
\usepackage{subfigure}
\usepackage[numbers, sort]{natbib}
\usepackage{color}
\usepackage{lipsum}
\usepackage{fancyhdr}

\newcommand{\partAbl}[2]{\frac{\partial #1}{\partial #2}}

\begin{document}

\title{Measuring and Comparing the Scaling Behaviour of a High-Performance CFD Code on Different Supercomputing Infrastructures}

\author{\IEEEauthorblockN{J\'{e}r\^{o}me Frisch}
\IEEEauthorblockA{Institute of Energy Efficiency\\ and Sustainable Building E3D\\
RWTH Aachen University\\
52074 Aachen, Germany\\
Email: frisch@e3d.rwth-aachen.de}
\and
\IEEEauthorblockN{Ralf-Peter Mundani}
\IEEEauthorblockA{Chair for Computation in Engineering\\
Technische Universit\"{a}t M\"{u}nchen\\
80333 M\"{u}nchen, Germany\\
Email: mundani@tum.de}
}

%


\maketitle


\begin{abstract}

Parallel code design is a challenging task especially when addressing petascale systems for massive parallel processing (MPP), i.\,e.\ parallel computations on several hundreds of thousands of cores. An in-house computational fluid dynamics code, developed by our group, was designed for such high-fidelity runs in order to exhibit excellent scalability values. Basis for this code is an adaptive hierarchical data structure together with an efficient communication and (numerical) computation scheme that supports MPP. For a detailled scalability analysis, we performed several experiments on two of Germany's national supercomputers up to 140,000 processes. In this paper, we will show the results of those experiments and discuss any bottlenecks that could be observed while solving engineering-based problems such as porous media flows or thermal comfort assessments for problem sizes up to several hundred billion degrees of freedom.

\end{abstract}


\pagestyle{empty}
\thispagestyle{fancy}
\lhead{}
\chead{}
\rhead{}
\lfoot{\copyright 2016 IEEE\\ 2015 17th International Symposium on Symbolic and Numeric Algorithms for Scientific Computing (SYNASC), Timisoara, 2015, pp. 371-378. doi: 10.1109/SYNASC.2015.63}
\cfoot{}
\rfoot{}

\flushbottom

%

\section{Introduction and Motivation}

Modern supercomputers tend to be massive parallel, i.\,e.\ they consist of several hundreds of thousands of cores, thus making efficient code design inevitable in order to exploit the underlying performance and to keep up with the so-called exascale challenge. While lot of research is currently happening into this direction, still plenty of codes are not prepared yet for scalability runs on more than 32,000 cores. In many cases, communication, i.\,e.\ data exchange between single processes, and load balancing -- considering adaptive mesh refinement, e.\,g.\ -- are the major problems, preventing such codes from high-fidelity computations on petascale systems such as SuperMUC, installed at the Leibniz Supercomputing Centre in Garching, with its more than 241,000 cores and a combined peak performance of 6.8 Petaflops or JuQueen, installed at J\"ulich Supercomputing Centre, with its more than 458,000 cores and an overall peak performance of 5.9 Petaflops.

On the other hand, the so-called emerging sciences such as medicine, sociology, biology, virology, chemistry, climate or geo-sciences demand for more and more computing power in order to solve multi-scale, multi-physics, and frequently also multi-domain problems. Those problems are not only complex by their nature, they often address further aspects such as big data or real time (in-situ) computing -- the latter one enabling decision makers to decide quickly where late results would be useless -- thus putting a lot of pressure on the parallel code design. What all these codes have in common is the necessity for efficient data structures, fast communication schemes, state-of-the-art numerical algorithms, and advanced load balancing techniques -- all in all key factors for the successful deployment of scalable massive parallel applications on modern petascale systems.

In this paper, we present a computational fluid dynamics (CFD) code for various applications such as thermal comfort assessment or porous media flows. Basic principle of this code is a hierarchical data structure in combination with an efficient multi-grid-like solver that perfectly scales both in the spatial (geometry) and computational (cores) domain. This structure allows to adaptively refine the computational mesh during runtime as well as to migrate blocks of the grid between processes in order to achieve an optimal load situation concerning data locality and minimal communication. A special process -- called neighbourhood server -- keeps track of all other (worker) processes, the data assigned to them, and the communication pattern among all nodes. Due to this neighbourhood server an efficient orchestration of the nodes becomes possible, thus we are able to obtain very good scalability and speed-up values when running the code on up to 140,000 processes on different supercomputers.

Main focus of this paper is a detailed scalability study of our CFD code on the two aforementioned petascale systems, i.\,e.\ SuperMUC and JuQueen, based on an indoor thermal comfort simulation. Therefore, we will profile typical communication properties along with special code characteristics, compare those results obtained on the two systems, and finally discuss identified bottlenecks and weak aspects of our parallelisation concept. These results do not only allow us to get a valuable insight into the code, but also to reveal those points which need further tuning in order to be ready for the next step towards the exascale challenge. The remainder of this paper is structured as follows: In the next section we will present the mathematical model used in our code, followed by a brief description of the data structure and the communication concept. Section~3 will highlight a scalability study performed on two different petascale systems together with a sound analysis of the measured results, while Section~4 addresses sample scenarios computed on several thousand cores using our CFD code. Section~5 will then close the paper with a short summary and outlook.

\section{MPFluid -- Massive Parallel CFD Code}
\label{sec:data_struct}

\subsection{Mathematical Modelling}

The mathematical background of the code is described in detail in Frisch et al.~\cite{Frisch2015Computation}. This section aims at giving a concise introduction into the mathematical modelling and into the data structure in order to bring the reader up to speed and to motivate the usage of HPC methods.

The mathematical modelling of the implementation is based on the Navier--Stokes equations derived from the conservation of mass, momentum, and energy principles. A complete and extensive derivation of the equations can be found in standard literature \cite{Batchelor2000, Ferziger2002, Hirsch2007}. The governing equations for an incompressible Newtonian fluid flow comprise three sets of equations. The first equation set -- given in differential form -- is called the continuity equation \vspace{-8pt}

\begin{equation}
\label{eq:cons_mass_incomp_div}
\nabla \cdot \vec u = 0 \quad ,\vspace{-2pt}
\end{equation}
where $\vec u$ describes the velocity of the fluid field using $u_1,u_2,u_3$ as velocity components in the three spatial directions $x_1,x_2,x_3$, respectively. This equation has to be satisfied at every time step in the complete domain. 

The second set, called the momentum equations, can be written for every direction $i \in \{1,2,3\}$ as \vspace{-6pt}

\begin{equation}
\label{eq:cons_mom_newtonian_fluid_incomp}
\partAbl{\rho_\infty u_i}{t} + \nabla \cdot (\rho_\infty u_i \vec u) = \nabla \cdot \left( \mu \nabla u_i \right) - \nabla \cdot \left(p \vec e_i \right) + b_i ~,
\vspace{6pt}
\end{equation}
where $t$ represents the time, $\rho_\infty$ the density of the fluid assumed constant over the complete domain and $\mu$ the dynamic viscosity. $p$ represents the pressure, and $b_i$ interior body forces in direction $i$. $\vec e_i$ represents the unit vector in direction $i$.

A temperature convection-diffusion equation can be applied additionally if thermal effects such as buoyancy should be modelled and describes the energy conservation and models the heat transport. The  Boussinesq approximation couples the convection-diffusion equation to the momentum equations as described in Lienhard and Lienhard~\cite{Lienhard2011} if some assumptions hold.

There is no independent equation for the pressure $p$ in Equations \eqref{eq:cons_mass_incomp_div} and \eqref{eq:cons_mom_newtonian_fluid_incomp}. The momentum equations include the pressure gradient $\nabla p$, whereas the incompressible continuity equation does not contain $p$ at all. By applying pressure correction methods as proposed by Harlow and Welch~\cite{HarlowWelch1965} in 1965 or the fractional step method (or projection method) introduced by Chorin~\cite{Chorin1967} in 1967, incompressible flows can be solved nevertheless. The methods are based on an iteration between velocity and pressure fields, where the pressure field acts in every step as a correction in order to fulfil the continuity equation for the velocity field. The pressure term $p$ in the incompressible equations is often referred to as `working pressure' as Equation \eqref{eq:cons_mom_newtonian_fluid_incomp} only contains the pressure gradient and not the absolute value itself.

By applying the fractional step method and choosing an~explicit Euler time discretisation for the temporal derivative $\partial/\partial t$, the momentum conservation equations in the direction $i$ for an intermediate time step $^\star$ can be written \vspace{-6pt}

\begin{equation}
\label{eq:cons_mom_newtonian_fluid_incomp_veloc_disc_intermed}
\rho_\infty \cdot \frac{u_i^\star - u_i^n}{\Delta t} = - \nabla \cdot (\rho_\infty u_i^n \vec u^{\,n}) + \nabla \cdot \left( \mu \nabla u_i^n \right) + b_i^n \quad \vspace{6pt}
\end{equation}
by neglecting the pressure gradient. $\Delta t$ denotes the time step size, the superscript $^n$ denotes the current time step $n$, and the superscript $^\star$ an intermediate time step between $n$ and $n+1$. The pressure term is now treated in a second step \vspace{-3pt}

\begin{equation}
\label{eq:cons_mom_newtonian_fluid_incomp_press_disc_intermed}
\rho_\infty \cdot \frac{u_i^{n+1} - u_i^\star}{\Delta t} = - \nabla \cdot \left(p^{n+1} \vec e_i \right) \quad .\vspace{4pt}
\end{equation}

The summation of the two Equations \eqref{eq:cons_mom_newtonian_fluid_incomp_veloc_disc_intermed} and \eqref{eq:cons_mom_newtonian_fluid_incomp_press_disc_intermed} results in the original Equation \eqref{eq:cons_mom_newtonian_fluid_incomp} with an explicit treatment of the velocity term (i.\,e.\ at time step $n$) and an implicit treatment of the pressure term (i.\,e.\ at time step $n+1$).

Using the divergence operator on Equation \eqref{eq:cons_mom_newtonian_fluid_incomp_press_disc_intermed} leads to \vspace{-2pt}

\begin{equation}
\label{eq:cons_mom_newtonian_fluid_incomp_press_disc_intermed_diveq}
\rho_\infty \cdot \frac{\nabla \cdot \vec u^{\,n+1} - \nabla \cdot \vec u^{\,\star}}{\Delta t} = - \Delta p^{n+1} \quad .\vspace{6pt}
\end{equation}

As the pressure term must lead to a~divergence-free velocity field at time step $n+1$ (fulfilment of the continuity Equation \eqref{eq:cons_mass_incomp_div}), the equation for determining the pressure at time step $n+1$ can be written as \vspace{-4pt}

\begin{equation}
\label{eq:cons_mom_newtonian_fluid_incomp_press_disc_poisson}
\Delta p^{n+1} = \frac{\rho_\infty}{\Delta t} \nabla \cdot \vec u^{\,\star} \quad ,\vspace{6pt}
\end{equation}
representing a Poisson equation for the pressure, which has to be solved in every time step.

The temporal discretisation can be enhanced using a second order explicit multi-level Adams--Bashforth method, introduced in detail by Schwarz~\cite{Schwarz2011}, for example. A very detailed analysis of other approaches for time derivatives can be found in Ferziger~\cite{Ferziger1998}. Kim and Moin~\cite{Kim1985} introduced in 1985 a~mixture of semi-implicit and explicit methods, where the convective terms in Equation \eqref{eq:cons_mom_newtonian_fluid_incomp_veloc_disc_intermed} are discretised using a~second order explicit Adams--Bashforth method. The viscous terms, however, are discretised using a~semi-implicit Crank--Nicolson method, which eliminates some numerical stability problems. Choi and Moin~\cite{ChoiMoin1994} used in 1994 a~similar approach, but integrate a~predictor-corrector method.

Similar to the temporal discretisation, a~spatial discretisation has to be introduced in order to describe and prepare the domain for a~numerical simulation. Popular, well-established methods, such as the finite difference method (FDM), the finite volume method (FVM) or the finite element method (FEM), could be used. This work focuses on the FVM and FDM approach for discretising the spatial domain and a~good description of these methods can be found in Ferziger and Peri\'c~\cite{Ferziger2002} or Hirsch~\cite{Hirsch2007}, for instance.

Numerical stability issues have to be taken into account as an explicit time discretisation method was chosen here. Courant and Friedrichs~\cite{CFL1928} analysed this problem in 1928 and proposed using an upwind-difference for the convective term and central differences for the diffusive term, and Peyret and Taylor~\cite{Peyret1983} give a good overview on a~detailed stability analysis for explicit FDM.

\subsection{Data Structure}

The data structure is based on block-structured, non-overlapping, orthogonal, regular, hierarchical grids. A topological grid management (called `l-grids') is responsible for all hierarchic information, such as parents and children relations, while data grids (called `d-grids') contain the actual data arrays, such as velocity and pressure fields, for example.

\begin{figure}[!ht]
	\centering
		\includegraphics[width=0.28\textwidth]{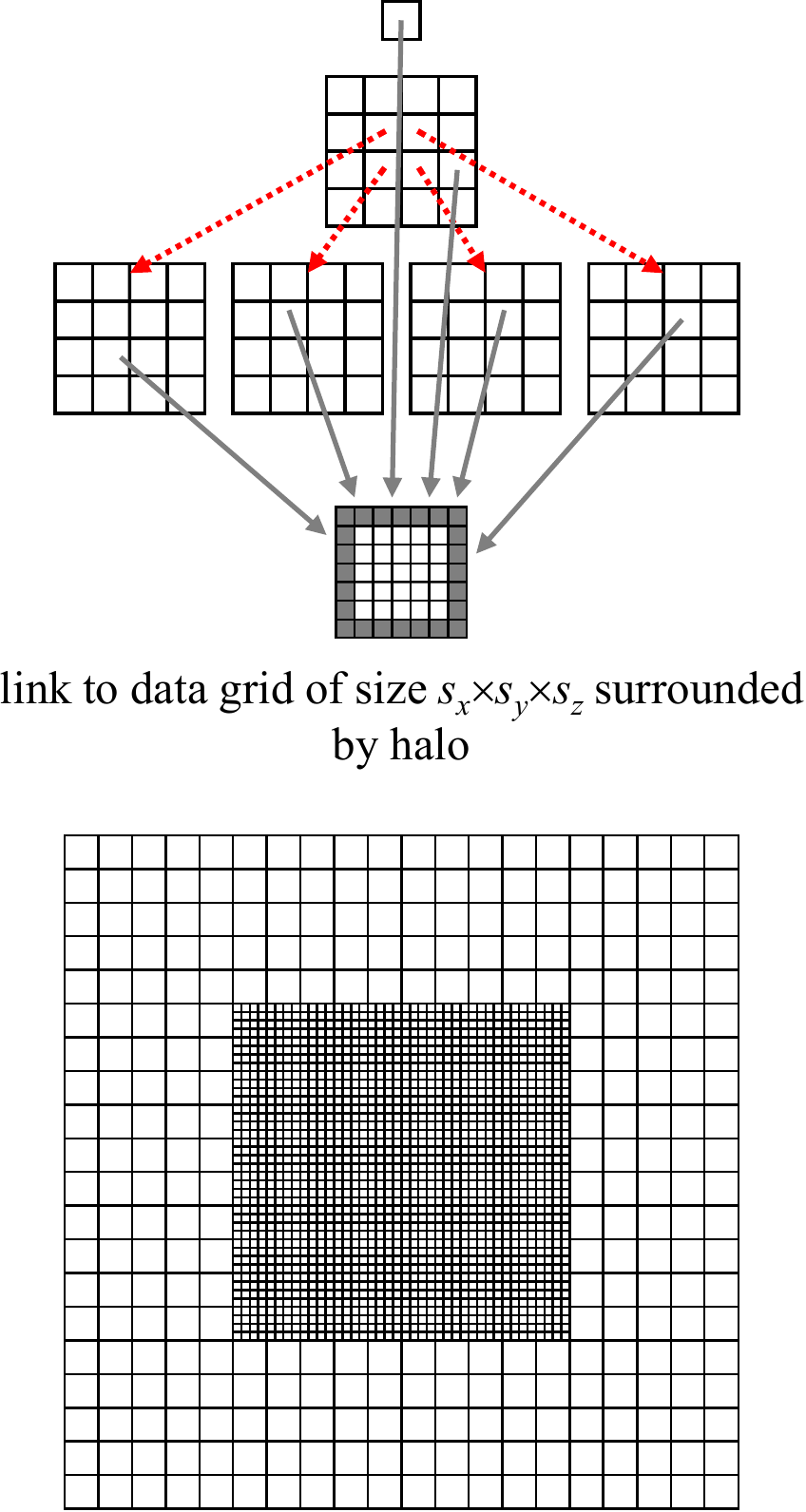}
	\caption{Schematic representation of the construction of the hierarchic grid data structure (based on Frisch~\cite{Frisch2014Diss}): Top part: logical hierarchy of non-overlapping adaptive grids. Bottom part: complete grid structure containing data grids.}
	\label{fig:data_structure}
\end{figure}

Figure~\ref{fig:data_structure} gives an overview of both data structure parts. In the top part, the nested construction of the non-overlapping grids can be seen. A root l-grid, located per definition at depth zero, is refined by $r_x^t,r_y^t,r_z^t$ in the respective directions. The new resulting child l-grids can be refined again by $r_x^s,r_y^s,r_z^s$ until the desired depth $d_{max}$ is reached. Furthermore, while creating the newly refined child l-grids, the coarse l-grid will be defined as their parent l-grid. Hence, each l-grid can only have one parent l-grid.

The second major part of the data structure consists of d-grids (i.\,e.\ the data grids). Every d-grid stores only the necessary variables, such as velocities, pressure, or temperature values in a matrix of cells. Thus, a cell is comparable to one control volume. The d-grid is surrounded by a halo of ghost cells necessary for the proper functioning of the structure. One of these d-grids can be seen in the middle part of Figure~\ref{fig:data_structure} directly above the text line and consists of the size of $s_x,s_y,s_z$, not counting the halo cells. Furthermore, each l-grid contains exactly one link to one d-grid. 

\vspace{-2pt}

Thus, complex, adaptive domain scenarios can be generated by combining regular blocks due to the flexible structure of the logical grid management. Each d-grid has an equidistant orthogonal spacing, and it can be shown that the finite volume approach using the mid-point rule and linear interpolation degenerates into a finite difference approach. Thus, on one d-grid, a simple finite difference scheme, such as a six-point stencil in 3D, can be used. This allows a strict separation into two phases: a computation phase and a communication phase. In the computation phase, the finite difference approach in the form of a stencil is evaluated on every d-grid. In the communication phase, a halo update is performed filling the ghost cells with neighbouring values in order to apply a Schwarz decomposition method~\cite{Schwarz1870}. In this phase the flux continuity has to be guaranteed on grid boundaries by the communication routines.

\subsection{Data Exchange and Neighbourhood Server}

As the data grids iterate solely over their local data, all other information, such as neighbouring and parental relations, must be provided by the logical grid management structure. Hence, the logical grid management takes care of flux conservation across d-grid boundaries, especially in the case of adaptive refinements, as depicted in Figure~\ref{fig:data_structure}. Therefore, complex data exchange mechanisms introduced in \cite{Frisch2011Synasc} are provided by the logical grid management framework to guarantee data integrity and consistency.

The distribution of grids to different processes is managed by a special dedicated process called `neighbourhood server'. It is described in detail in Frisch~\cite{Frisch2014Diss}. A dedicated server keeps track of the logical structure of all grids without knowing values of the data content itself and organises the distribution to different processes according to a Lebesgue space-filling curve (also called Z-order curve, see Bader \cite{Bader2013}) in order to preserve neighbouring relations. Performance measurements for exchange times of halo values show very good results in terms of grid-to-grid communication for the given distribution on high-performance computers and are presented in Section \ref{sec:scaling}.

\subsection{Multi-Grid-Like Pressure Poisson Solver}

As indicated in Equation \eqref{eq:cons_mom_newtonian_fluid_incomp_press_disc_poisson}, the fractional step method leads to a Poisson equation for the pressure term which has to be solved at every time step. Frisch \cite{Frisch2014Diss} shows that more than 90\,\% of the time is spend in the solution of the Poisson equation. Hence, the efficient solution of the pressure Poisson equation is of utmost importance.

In the 1970s Brandt~\cite{Brandt1977} introduced geometric multi-grid solvers as multi-level methods which represent an adequate technique for solving partial differential equations of elliptic type, such as the Poisson equation or the Laplacian equation. By comparing the data flows of restriction and prolongation operators to the already implemented data exchanges, many similarities can be seen and a cell-centred, multi-grid-like solver was applied on the already implemented communication structures.

In the following section, scaling results for different machines are compared for a Laplacian equation, as this resembles the pressure Poisson equation. The Laplacian problem \vspace{-4pt}

\begin{equation}
\label{eq:adap_prob_setup}
\partAbl{^2 p(x,y,z)}{x^2} + \partAbl{^2 p(x,y,z)}{y^2} + \partAbl{^2 p(x,y,z)}{z^2} = 0 \vspace{6pt}
\end{equation}
is defined on a~1\,$\times$\,1\,$\times$\,1\,m cubic domain, where fixed Dirichlet boundary conditions are applied on the east and west side according to \vspace{-8pt}

\begin{equation}
\label{eq:adap_prob_setup_bc_east_west}
p(0,y,z) = p(1,y,z) = 1 \qquad \forall~ y,z \quad ,
\end{equation}
whereas all other sides are set to $p=0$. This equation describes a diffusion process and can be used as well for modelling a stationary temperature diffusion problem without any convective influences or internal loads.

\section{Scaling Results}
\label{sec:scaling}

In the following section, performance measurements are shown for tests on different machines for several depths of the hierarchical data structure. The measurements were performed on three supercomputing systems: Shaheen, a~16-rack, 65,536 cores IBM Blue~Gene/P supercomputer installed at King Abdullah University of Science and Technology (KAUST) in the Kingdom of Saudi Arabia, JuQueen, a 28-rack, 458,752 cores IBM Blue~Gene/Q supercomputer installed at J\"ulich Supercomputing Centre (JSC) in J\"ulich, Germany, and SuperMUC, a 155,656 core (before the hardware upgrade in 2015) large IBM System iDataPlex supercomputer installed at the Leibniz Supercomputing Centre (LRZ) in Garching near Munich, Germany.

All following measurements were done solely for the pressure Poisson Equation \eqref{eq:cons_mom_newtonian_fluid_incomp_press_disc_poisson} which represents the most computational complex part of the CFD code. This can be done without restriction of any kind, as the overhead for a full time step is according to \cite{Frisch2014Diss} only slightly larger than for just solving the pressure Poisson equation.

First of all, we measured the time for one total ghost layer exchange based on a fully refined 3D domain using a refinement level of (2,2,2) for the l-grids and a d-grid size of (16,16,16), thus resulting in computational blocks of 4,096 d-grid cells each. This ensemble of l-grid and d-grid sizes has proven to be optimal for SuperMUC, hence we were using this setup throughout all experiments on all systems. The measurements were then performed for different depths of the logical grids, ranging from 5 to 8, leading to approximately 78.5 billion d-grid cells with 9 independent variables per cell, i.\,e.\ more than 707 billion variables to be exchanged across all processes on depth 8.

\begin{figure}[!htbp]
	\centering
	\subfigure[][time plotted versus the number of processes used for computing]{%
\label{fig:impl:ghost_layer_exchange_times:per_process}
\includegraphics[height=8.3cm, angle=-90]{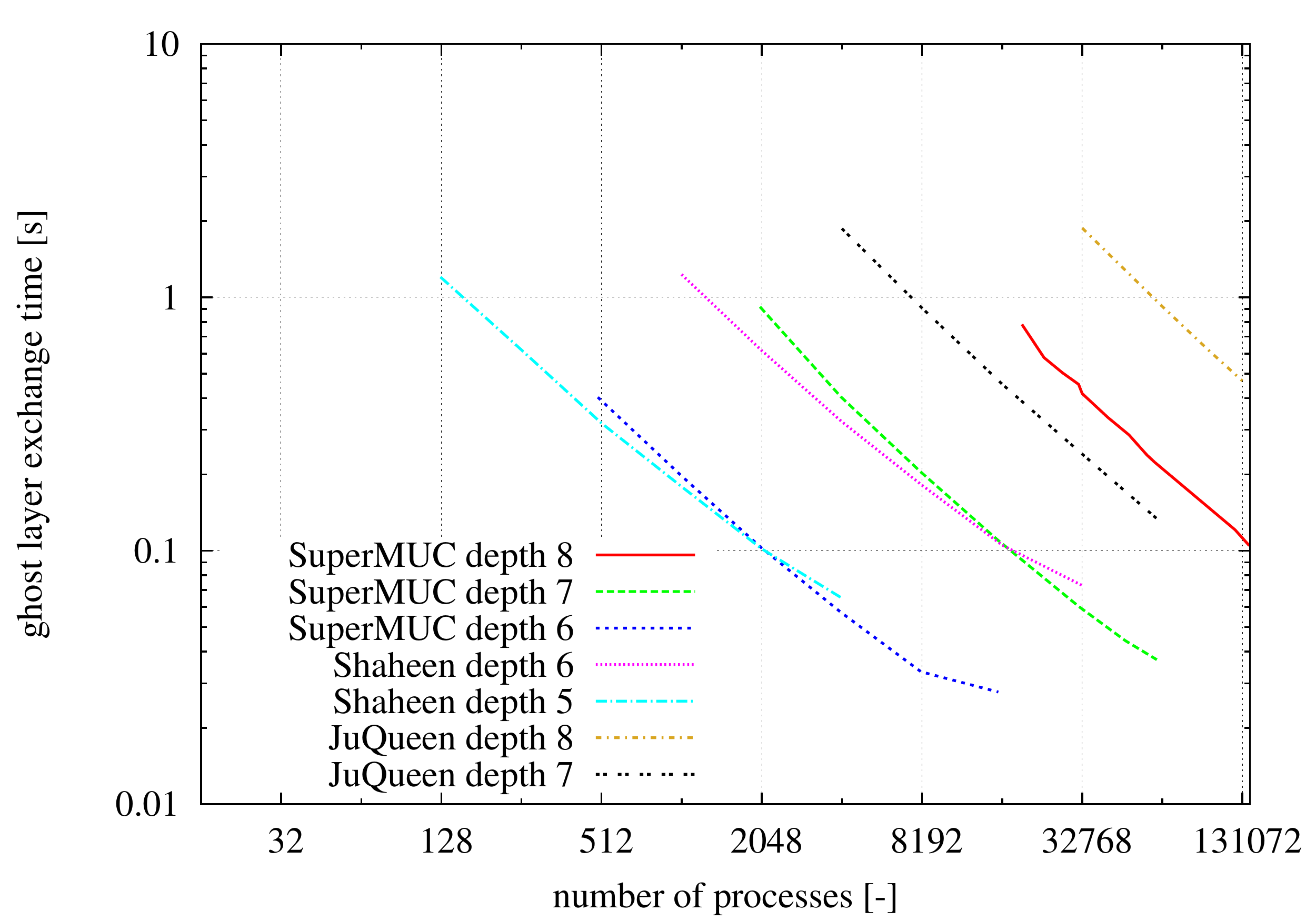} } \\
	\subfigure[][time plotted versus the number of l-grids per process]{%
\label{fig:impl:ghost_layer_exchange_times:per_grid_per_process}
\includegraphics[height=8.3cm, angle=-90]{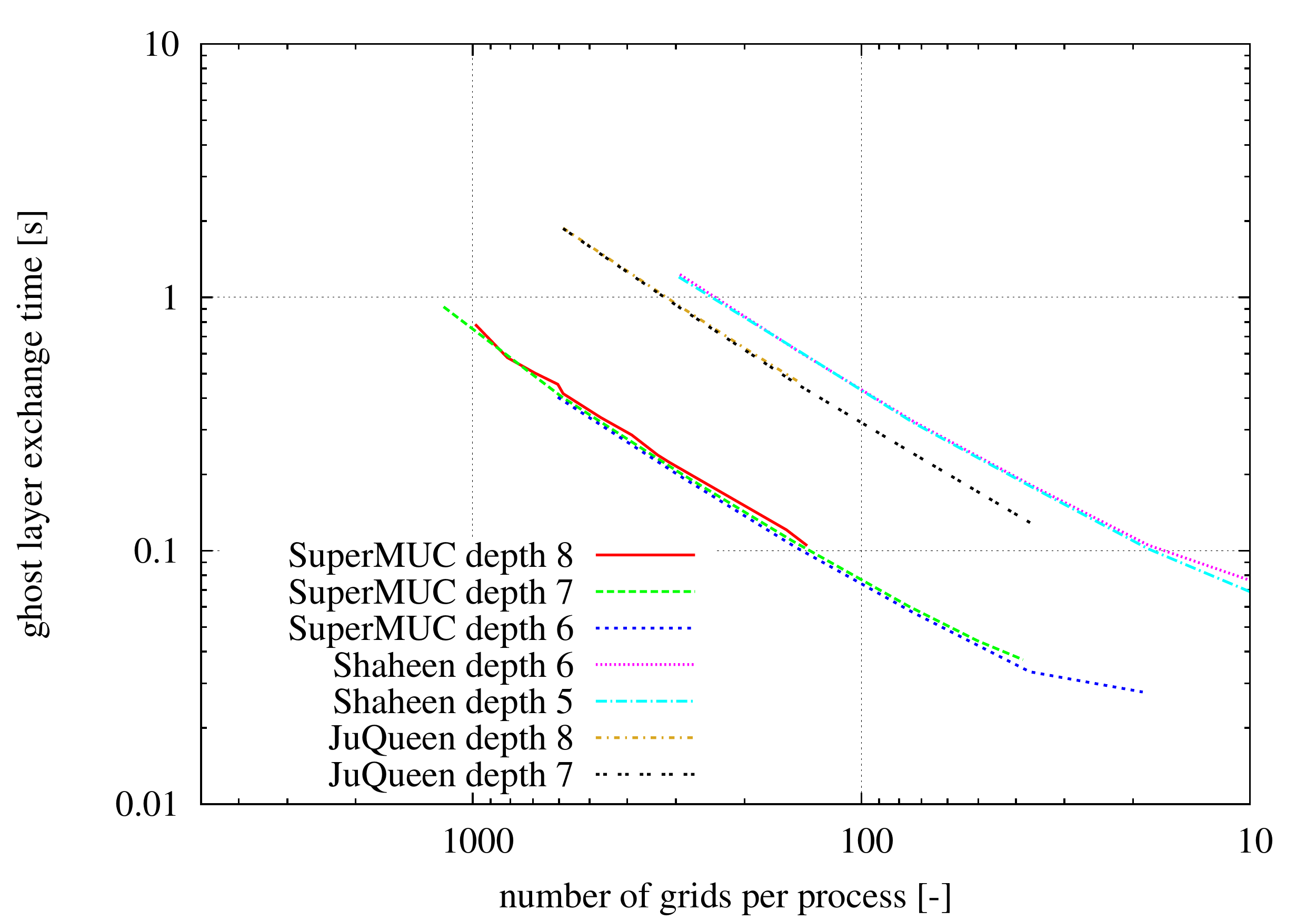} }
	\caption{Communication times for a total ghost layer exchange of 9 independent variables per cell in seconds on three different computing platforms (SuperMUC, JuQueen, Shaheen) for a 3D domain of fully refined l-grids with a refinement level (2,2,2) up to depth 8 and a d-grid size of (16,16,16). (Figure (b) is plotted in reverse order in order to indicate the increasing number of processes with decreasing number of grids per process for a given constant problem size.)}
	\label{fig:impl:ghost_layer_exchange_times}
\end{figure}

In Figure~\ref{fig:impl:ghost_layer_exchange_times} (a), the total exchange times in [s] are depicted for the three different systems. All curves show a clear tendency to decrease for an increasing number of processes. On SuperMUC, for instance, the total exchange time between more than 140,000 processes on depth 8 with over 707 billion variables is approximately 0.1\,s and, thus, practically negligible during the computation. Interestingly, the curves on SuperMUC and the two IBM Blue~Gene systems not only look very similar, but also exhibit the same slope -- except the fact that the exchange times on SuperMUC are faster than on the Blue Gene systems due to SuperMUC's higher clock frequency -- which is already a first indication of the data structure's well scalability characteristics and its suitability for massive parallel computations.

In order to relativise the influence of the different clock frequencies, in Figure~\ref{fig:impl:ghost_layer_exchange_times} (b) the total exchange times in [s] are plotted against the number of l-grids per process for the three different systems. It is also observable that a hardware upgrade from Blue~Gene/P to Blue~Gene/Q had an impact on the communication times. The slopes are still similar but show an offset between both machines. One can clearly observe now the perfect accordance of all measurements for corresponding depth (where available) which further emphasises the sound scalability of our CFD code for different problem sizes, different amount of processes, and different architectures.

\begin{figure}[!htbp]
	\centering
	\subfigure[][time plotted versus the number of processes used for computing]{%
\label{fig:impl:time_to_solution:per_process}
\includegraphics[height=8.3cm, angle=-90]{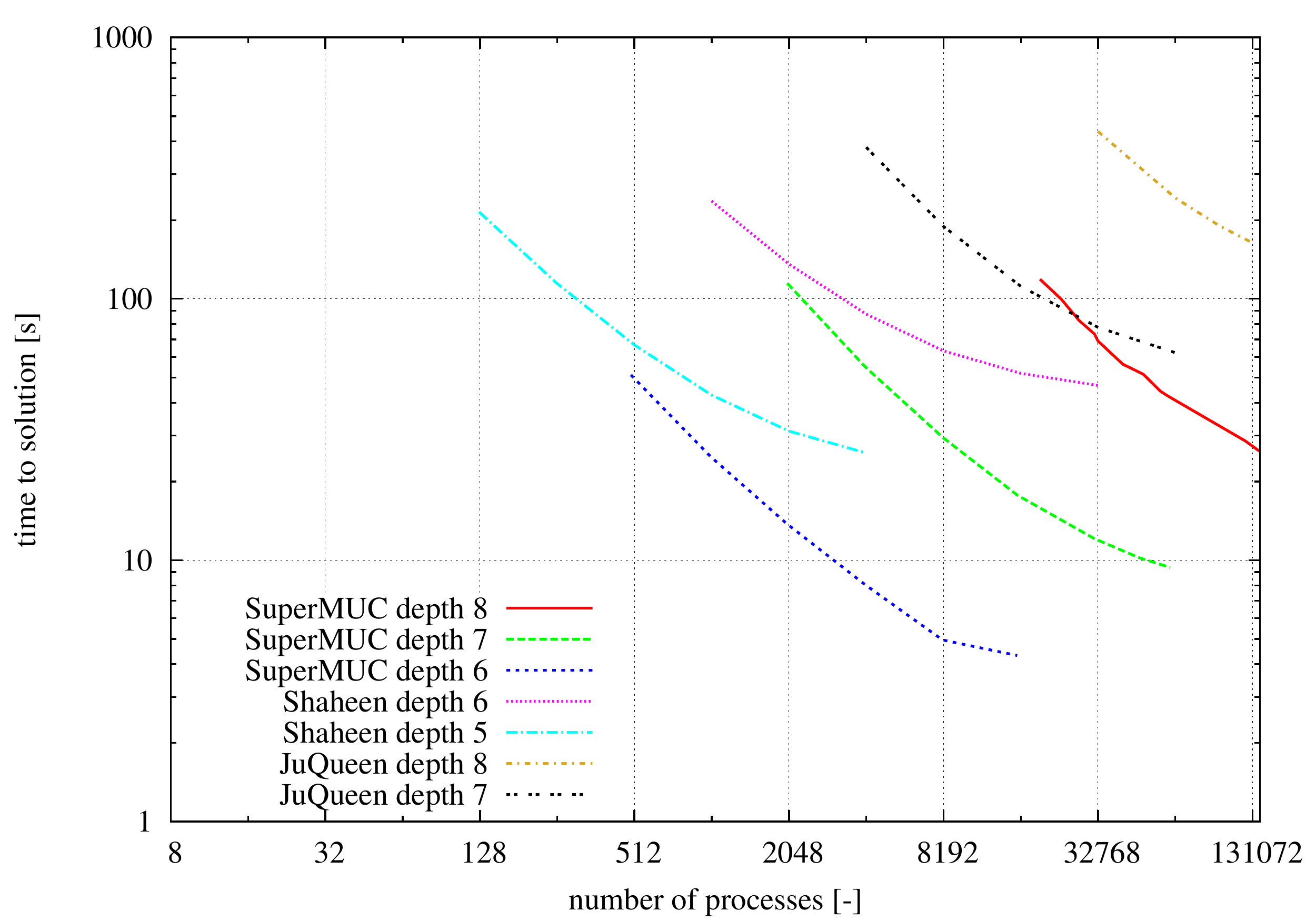} } \\
	\subfigure[][time plotted versus the number of l-grids per process]{%
\label{fig:impl:time_to_solution:per_grid_per_process}
\includegraphics[height=8.3cm, angle=-90]{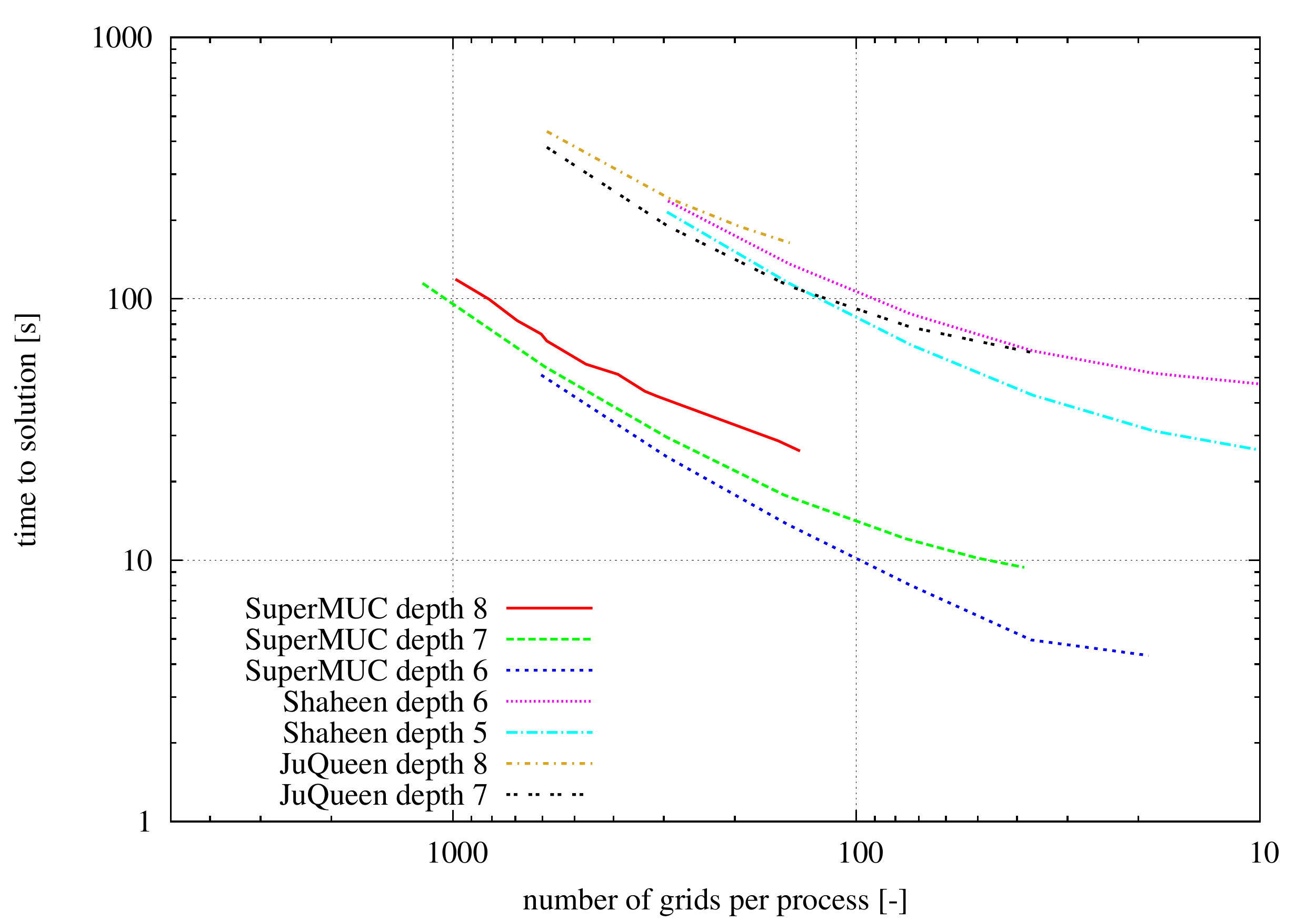} }
	\caption{Time to solution on three different computing platforms (SuperMUC, JuQueen, Shaheen) for a 3D domain of fully refined l-grids with a refinement level (2,2,2) up to depth 5, 6, 7, or 8 and a d-grid size of (16,16,16). (Figure (b) is plotted in reverse order in order to indicate the increasing number of processes with decreasing number of grids per process for a given constant problem size.)}
	\label{fig:impl:time_to_solution}
\end{figure}

Next experiments addressed a full time step update, i.\,e.\ an iterative solution of the pressure Poisson equation until convergence was reached. The setup was exactly the same as in the previous experiment, namely a fully refined 3D domain using l-grid refinement levels of (2,2,2) and d-grid sizes of (16,16,16) for different depths (5--8), leading to around 20 million l-grids with more than 78.5 billion d-grid cells and over 707 billion variables on depth 8. It is worth notifying that this setup also requires 28~TByte of combined main memory for storing all relevant data. In order to face this problem on SuperMUC, for instance, a total of at least 20,000 processes is necessary due to SuperMUC's memory availability of 1.5~GByte per core. On JuQueen, however, at least 32,000 processes were necessary to accommodate all relevant data due to a smaller memory per core ratio. In Figure~\ref{fig:impl:time_to_solution}, the times to solution (i.\,e.\ a full time step) in [s] are plotted against the total number of processes (a) and the total  number of l-grids per process (b) for the three different systems. Again, a clear decreasing tendency of all curves for an increasing number of processes is to be observed. Furthermore, the curves of SuperMUC and JuQueen still show the same slope (upper plot of Figure~\ref{fig:impl:time_to_solution}) and are more or less shifted in vertical direction only due to the two systems' different clock frequencies. The lower plot of Figure~\ref{fig:impl:time_to_solution} reveals a similar good accordance of all corresponding experiments on the different systems as in the previous case of a pure ghost layer exchange, negelecting hardware specific influences such as processor speed. Hence, the supposed scalability of the code could be approved by the second measurements, too.

\begin{figure}[!htbp]
	\centering
	\subfigure[][strong speedup from 1 to 8,192 processes]{%
\label{fig:impl:strong_speedup:8k}
\includegraphics[height=8.3cm, angle=-90]{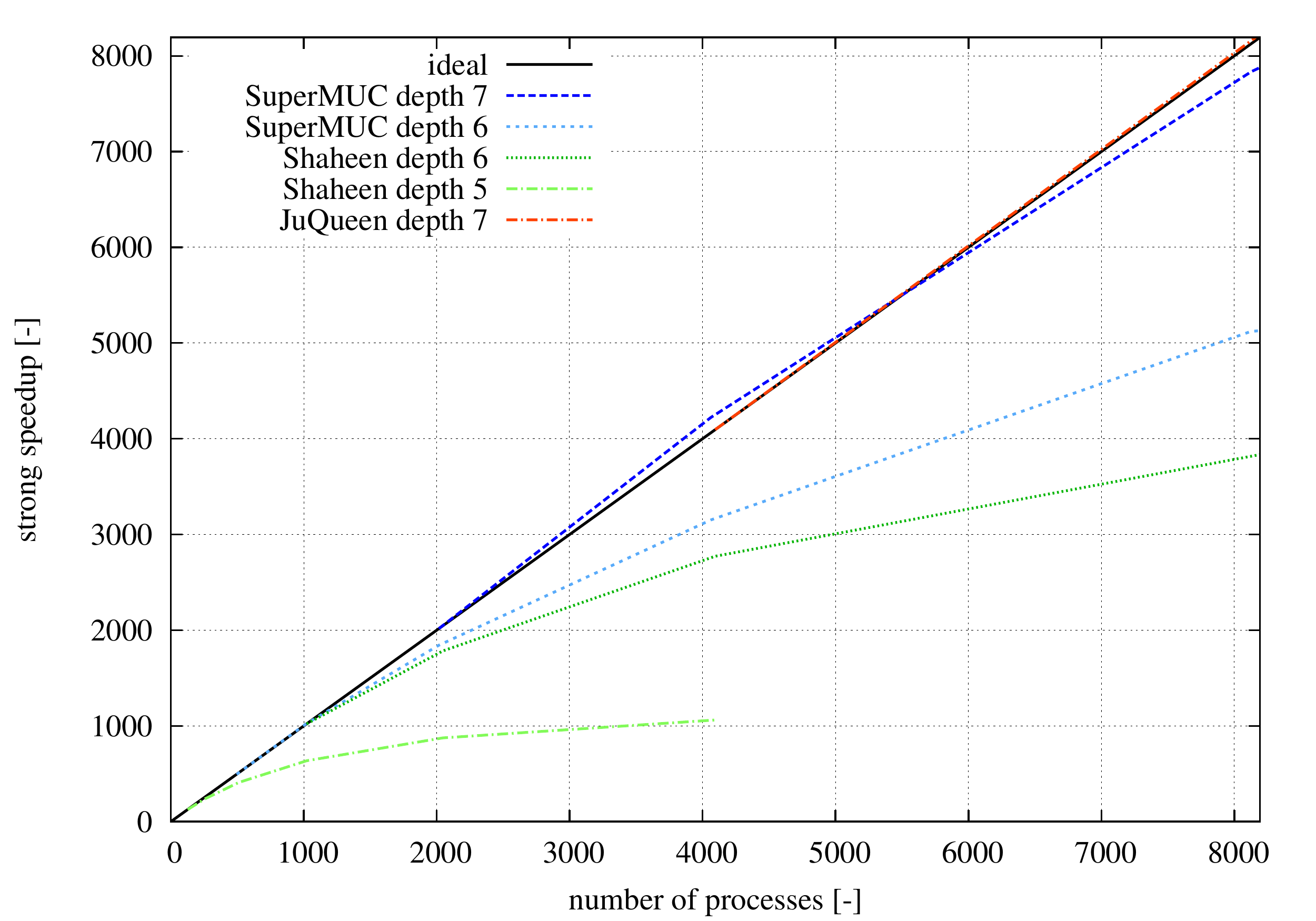} } \\
	\subfigure[][strong speedup from 1 to 16,384 processes]{%
\label{fig:impl:strong_speedup:16k}
\includegraphics[height=8.3cm, angle=-90]{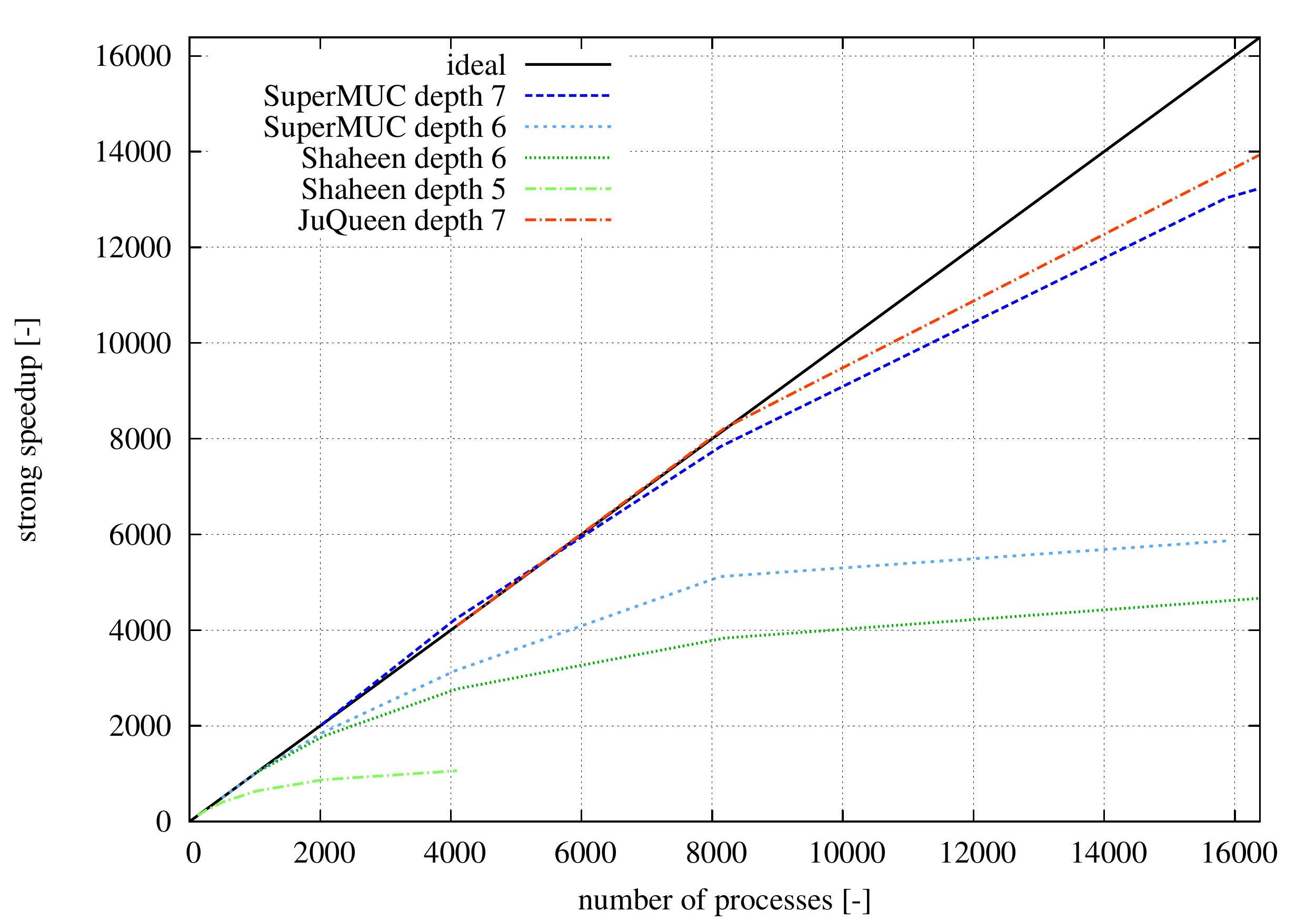} } \\
	\subfigure[][strong speedup from 1 to 32,768 processes]{%
\label{fig:impl:strong_speedup:32k}
\includegraphics[height=8.3cm, angle=-90]{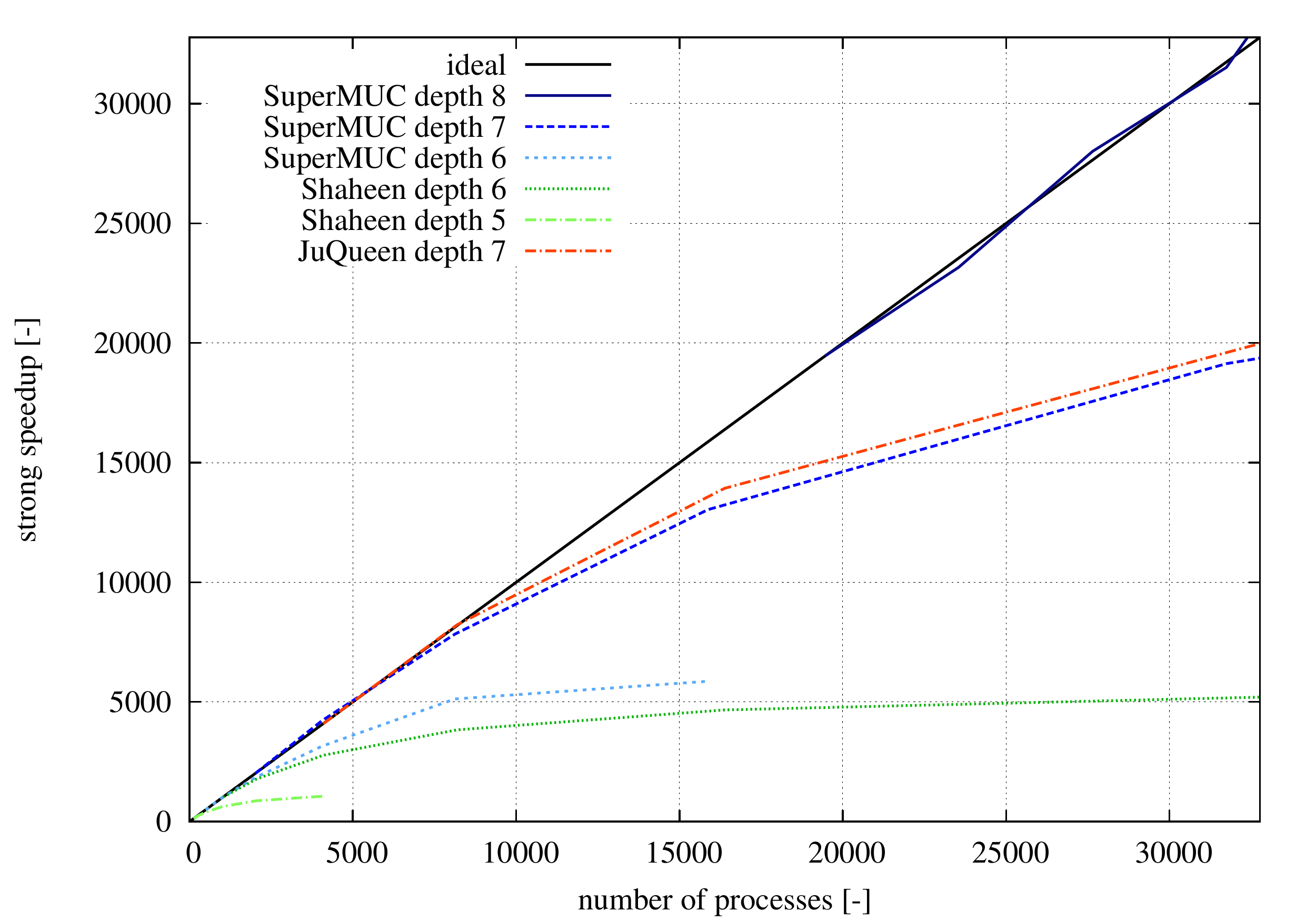} }
	\caption{Strong speedup results on three supercomputing platforms (SuperMUC, JuQueen, Shaheen) for a 3D domain of fully refined l-grids with a refinement level (2,2,2) up to depth 5, 6, 7, or 8 and a d-grid size of (16,16,16). The different speedup results are zoomed snapshots around the area of interest.}
	\label{fig:impl:speedup1}
\end{figure}

\begin{figure}[!htbp]
	\centering
	\subfigure[][strong speedup from 1 to 65,536 processes]{%
\label{fig:impl:strong_speedup:64k}
\includegraphics[height=8.3cm, angle=-90]{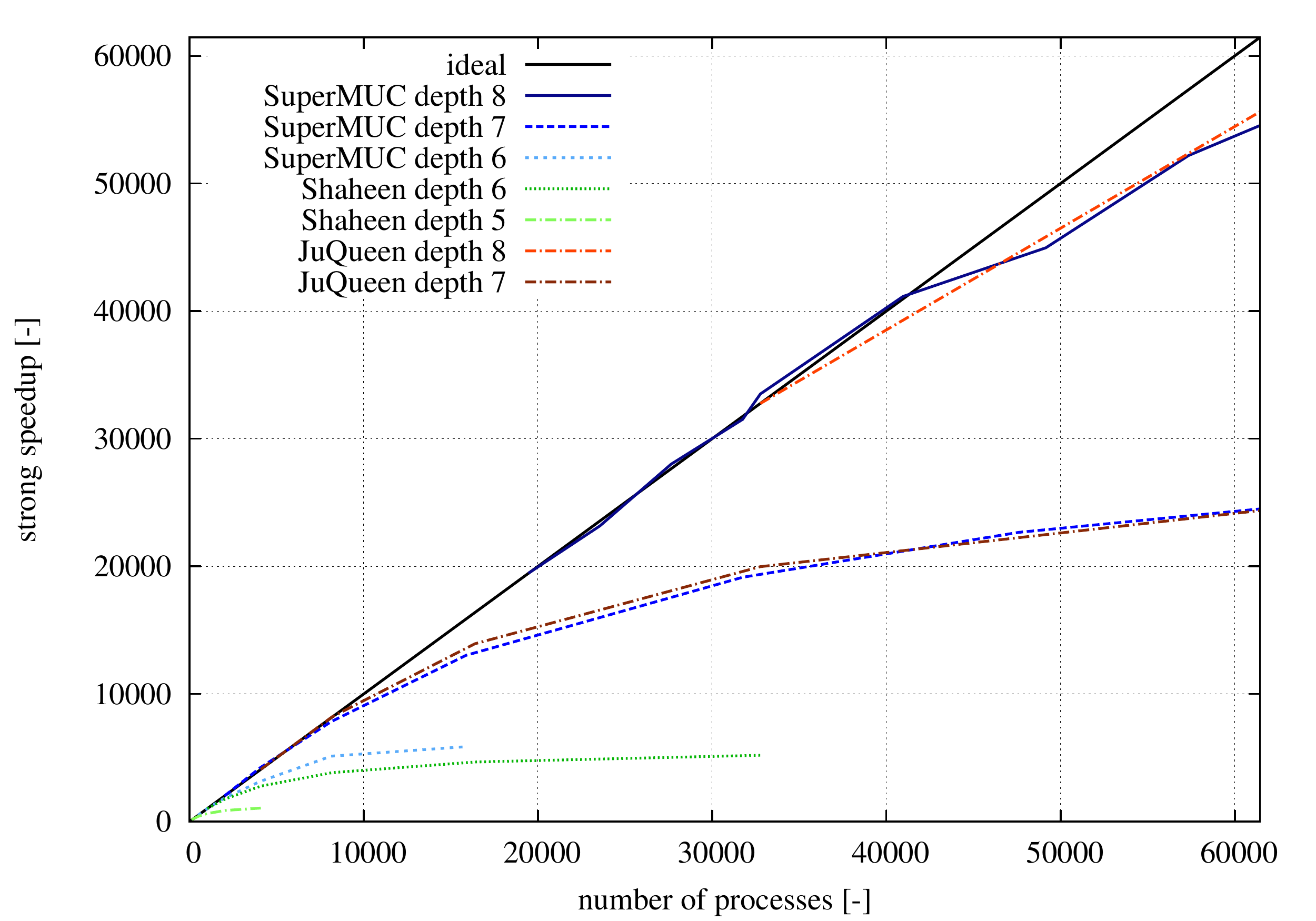} } \\
	\subfigure[][strong speedup from 1 to 139,008 processes]{%
\label{fig:impl:strong_speedup:139k}
\includegraphics[height=8.3cm, angle=-90]{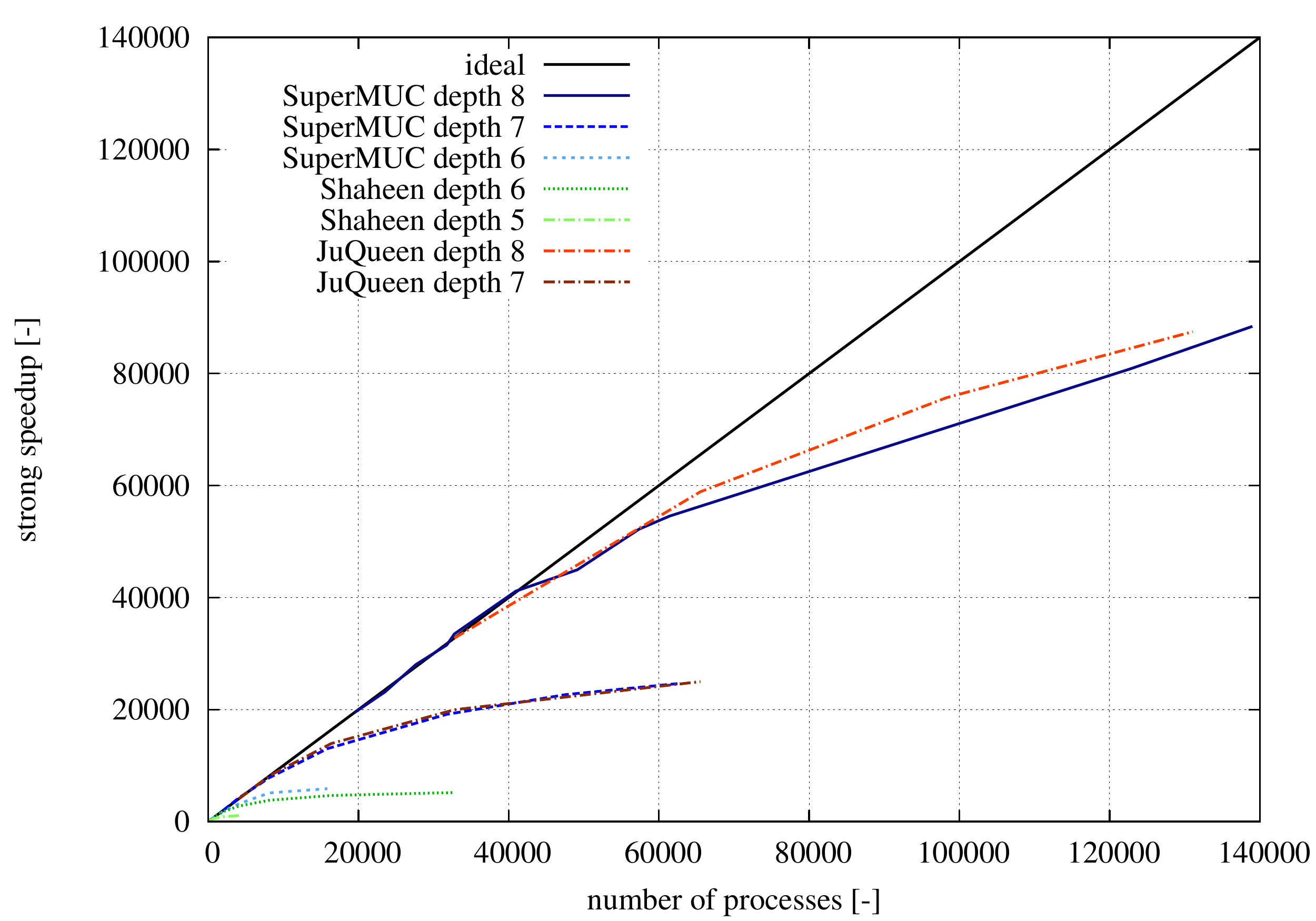} }
	\caption{Strong speedup results on three supercomputing platforms (SuperMUC, JuQueen, Shaheen) for a 3D domain of fully refined l-grids with a refinement level (2,2,2) up to depth 5, 6, 7, or 8 and a d-grid size of (16,16,16). The different speedup results are zoomed snapshots around the area of interest.}
		\label{fig:impl:speedup2}
\end{figure}

Another interesting aspect is the maximum throughput for a given (fixed) problem size and different amounts of processes -- also known as strong speed-up. Figures~\ref{fig:impl:speedup1} and \ref{fig:impl:speedup2} highlight those strong speed-up measurements for the three different systems with up to a total of 8,192, 16,384, 32,768, 65,536, and 139,008 processes. The experimental setup is the same as in the previous cases. For better visibility, all of the above five cases are plotted separately against the ideal speed-up with a slope of 1.

In Figure~\ref{fig:impl:speedup1}, one general observation is that all curves start to level-off as soon as the underlying problem becomes too small and communication time (between cores and nodes) dominates computation time. In a direct comparison between SuperMUC and Shaheen, i.\,e.\ Intel's Sandy-Bridge architecture vs.\ IBM's PowerPC 450 generation, both CPUs have similar last-level (L3) cache sizes while the Sandy-Bridge architecture has twice as much cores. Here, we can observe better (up to a factor of 1.5) performance values for the Intel architecture based on the same problem size, indicating that currently the code is not memory-bound, i.\,e.\ limited by the memory interface. In comparison to the Blue~Gene/Q system (JuQueen) with its four times larger, 16 way set-associative last-level (L2) cache and its four-times more cores the obtained results look very identical to those of SuperMUC. Even one would expect much better results for JuQueen, the two systems achieve a very similar performance which -- again -- underlines that the code does not suffer from a memory-bound limitation at the moment. Hence, any performance loss must come from the computational kernel that does not exploit intrinsic features such as SMT, SIMD, or vectorisation (pipelining) yet. A detailed analysis according to the roofline model \cite{Roofline2009} is inevitable in order to perform code optimisations for a higher node-level performance.

Finally, Figure~\ref{fig:impl:speedup1} shows the obtained strong speed-up values for SuperMUC and JuQueen on 65,536 and 139,008 processes. The different curves of the two systems for different problem sizes once more perfectly coincide and lead to a parallel efficiency of 64\,\% on depth 8. While the levelling-off of both systems was to be expected (being a well-known feature of such analyses), the identical performance of both systems gives rise to further considerations. On the one hand, we observed a good scalability of our code w.\,r.\,t.\ to the communication, i.\,e.\ to the ghost layer exchange, on the other hand, any performance benefit of the Blue~Gene/Q cannot be utilised by the computational kernel in order to outperform the Sandy-Bridge architecture. In consequence the data structure has proven to suffice very well for massive parallel computations, whereas the computational kernel turns out to be currently a bottleneck hindering the exploitation of the underlying performance. Further experiments with adaptive grids -- that do not follow a uniform refinement of the computational domain -- revealed very similar results and, thus, led to the same conclusions. More information on those measurements can be found in \cite{Frisch2014Diss}.

\section{Application Examples}

In order to show the versatility of the proposed approach, a few application examples are listed below.

\begin{figure}[ht]
	\centering
		\includegraphics[width=0.48\textwidth]{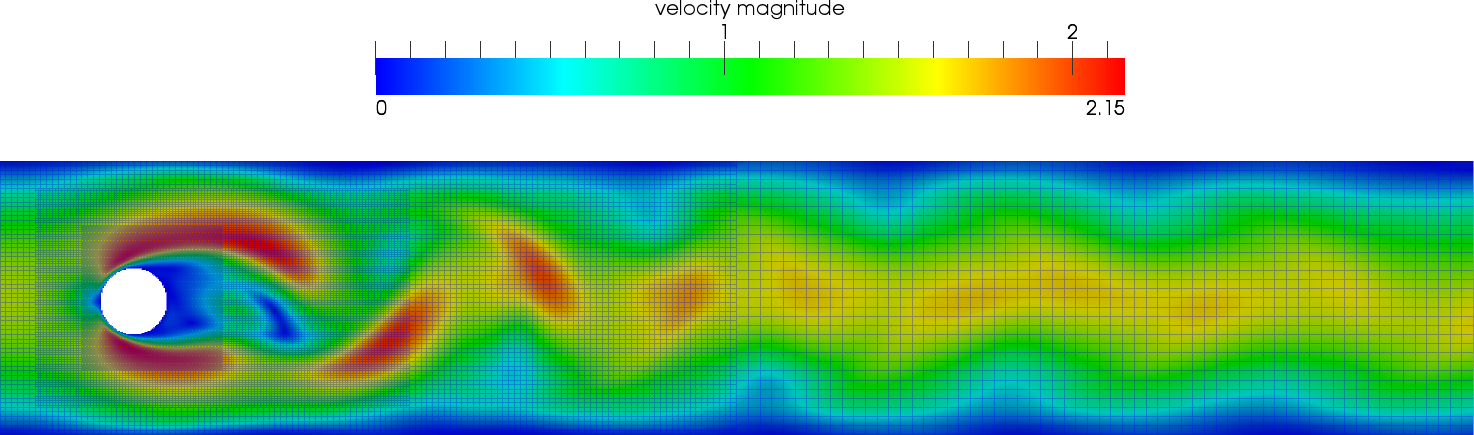}
	\caption{Simulation of a von K\'arm\'an vortex street according to the Sch\"afer-Turek \cite{SchaeferTurek1996} benchmark 2D-2 with $Re=100$ showing the well-known vortex shedding.}
	\label{fig:schaefer_tureg_2D-2}
\end{figure}

Figure \ref{fig:schaefer_tureg_2D-2} shows a typical benchmark scenario for validating CFD codes. The DFG priority research program `Flow Simulation on High-Performance Computers' \cite{SchaeferTurek1996} introduces a collection of benchmarks. The basic setup consists of a regular rectangular channel in 2D or 3D with a cylindrical or rectangular obstacle in the inflow region of the channel. Here, the channel was computed using an adaptive grid setup. This example, based on the 2D-2 setup with $Re=100$, shows vortex shedding and was used in Frisch \cite{Frisch2014Diss} as one of the validation test cases.

\begin{figure}[ht]
	\centering
		\includegraphics[width=0.30\textwidth]{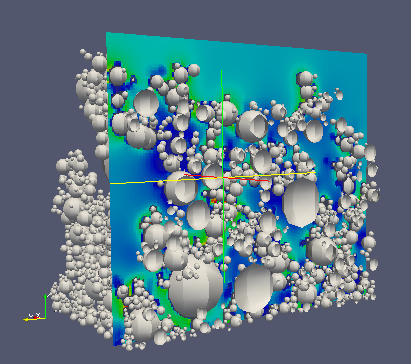}
	\caption{Simulation of fluid flow through porous media (from \cite{Perovic2014}).}
	\label{fig:pic_porous_media}
\end{figure}

Figure \ref{fig:pic_porous_media} shows a flow through porous media. In Perovi\'c et al. \cite{Perovic2014}, the presented code was used for computing the micro-scale of fluid flow through porous media while coupling it on a macro-scale to a Darcy-based flow solver. Hence, this shows that the code can work with complex geometries without a time consuming manual meshing process, as the geometry generation is done fully-automatically by a voxel-based approach, thus generating a representation directly usable by the code's data structure.

\begin{figure}[ht]
	\centering
		\includegraphics[width=0.43\textwidth]{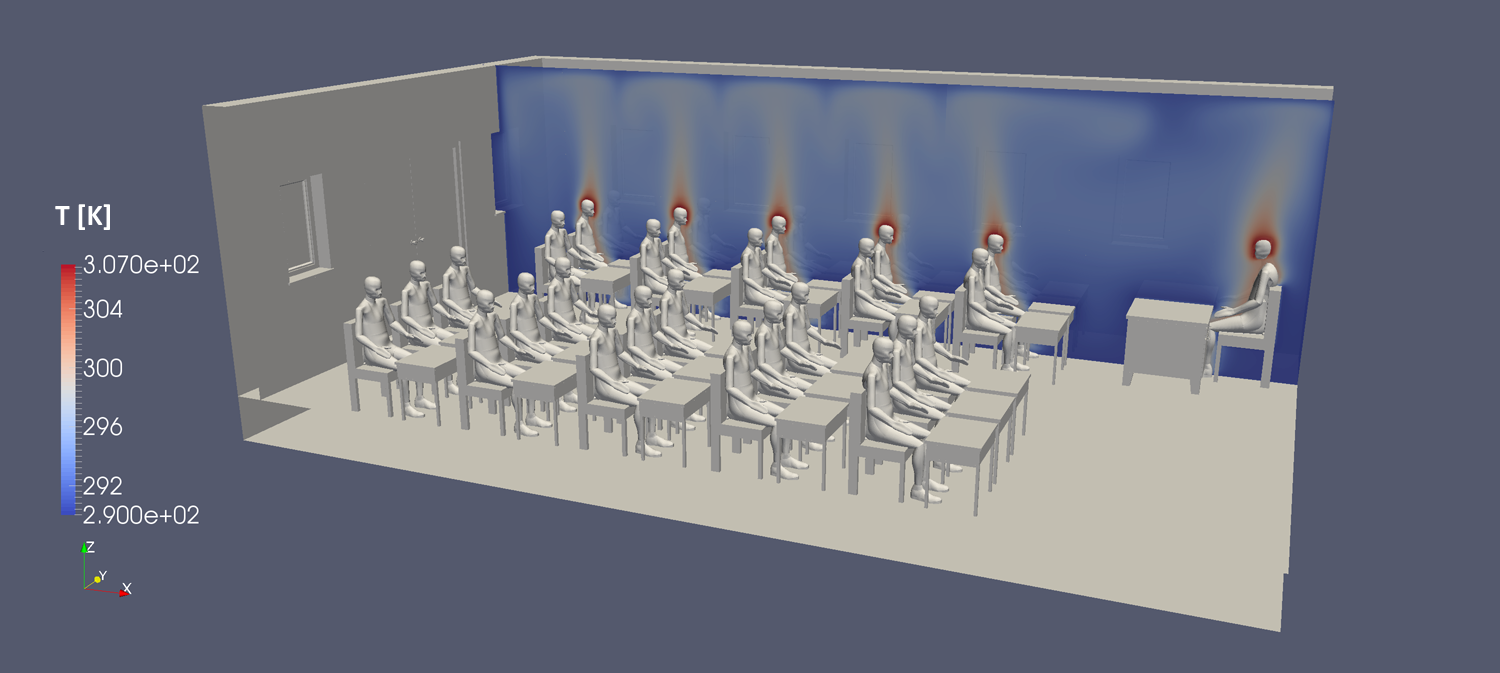}
	\caption{Simulation of thermally coupled fluid flow behaviour under natural convection boundary conditions in a classroom setup (from \cite{Frisch2015Computation}).}
	\label{fig:classroom_natural_ventilation}
\end{figure}

Last but not least, the code is also able to compute thermally coupled scenarios by applying the Boussinesq approximation as mentioned in Section \ref{sec:data_struct}. In this case, the flow is subjected to pure natural convection boundary conditions, meaning that no inflow or outflow conditions are set in the room and the flow is only driven by thermal buoyancy effects. Figure \ref{fig:classroom_natural_ventilation} shows a classroom setup, where the human occupants are coupled to a thermoregulation model imposing thermal boundary conditions onto the surfaces, and acting as driving forces for the natural convection scenario. Further information and result evaluation can be found in Frisch et al. \cite{Frisch2015Computation}.

Thus, the code is able to handle quite different physical scenarios on different scales while running on more than 100,000 cores.

\section{Conclusion}

In this paper, we have presented a massive parallel CFD code based on hierarchical data structures for the usage on modern petascale supercomputing systems. As this code should serve for various complex engineering-based application scenarios demanding for extensive computing power, scalability was an important design issue from the very beginning. Within several measurements performend on two of Germany's national supercomputers (both among the first 20 places in the current top 500 list\footnote{http://www.top500.org as of June 2015}) very good scalability characteristics up to 140,000 processes could be observed. Moreover, within those experiments it could be shown that any performance limitation of the computational kernel is not memory-bound and, thus, further performance improvement due to node-level optimisation is possible. Such optimisations comprise simultaneous multithreading (SMT) or streaming computations as well as the exploitation of intrinsic characteristics like the Advanced Vector Extensions (AVX) of Intel's Sandy-Bridge architecture in order to increase throughput capabilities.

\section{Acknowledgement} 

The authors gratefully acknowledge the computing time granted by the JARA-HPC Vergabegremium and provided on the JARA-HPC Partition part of the supercomputer JUQUEEN \cite{Juqueen2015} at Forschungszentrum J\"ulich.

Furthermore, the authors would like to cordially thank Leibniz Supercomputing Centre (LRZ) in Garching for the computing time granted during the `Extreme Scaling Workshop' in June 2014.


\bibliographystyle{IEEEbib}
\bibliography{paper}

\end{document}